\pdfoutput=1
\documentclass{article}
\usepackage{spconf,amsmath,graphicx,float}
\usepackage{multirow}
 \usepackage[compact]{titlesec}
\usepackage{newtxtext,newtxmath}
\usepackage{csquotes} 
\usepackage{longtable}
\usepackage{adjustbox}
\usepackage{graphicx}
\usepackage{graphics,xcolor}
\usepackage[makeroom]{cancel}
\usepackage[ruled,vlined]{algorithm2e}

\title{A Weighted Generalized Coherence Approach for Sensing Matrix Design}
\name{Ameya Anjarlekar, Ajit Rajwade}
\address{Indian Institute of Technology, Bombay}
\begin{document}
%
\maketitle
\begin{abstract}
As compared to using randomly generated sensing matrices, optimizing the sensing matrix w.r.t. a carefully designed criterion is known to lead to better quality signal recovery given a set of compressive measurements. In this paper, we propose generalizations of the well-known mutual coherence criterion for optimizing sensing matrices starting from random initial conditions. We term these generalizations as bi-coherence or tri-coherence and they are based on a criterion that discourages any one column of the sensing matrix from being close to a sparse linear combination of other columns. We also incorporate training data to further improve the sensing matrices through weighted coherence, weighted bi-coherence, or weighted tri-coherence criteria, which assign weights to sensing matrix columns as per their importance. An algorithm is also presented to solve the optimization problems. Finally, the effectiveness of the proposed algorithm is demonstrated through empirical results.
\end{abstract}
\section{Introduction}
\label{sec:intro}
Compressed sensing is a signal processing technique for efficiently acquiring and reconstructing a signal $\boldsymbol{x} \in \mathbb{R}^n$, by finding solutions to under-determined linear systems \cite{Candes_2008_ric}. For reconstruction, the sparsity or compressibility of signals in a dictionary $\boldsymbol{\Psi}$ (typically a discrete cosine transform (DCT) or a wavelet transform) is utilized. Such signals can be recovered from $m \ll n$ (noisy) linear projections of the form $\boldsymbol{y} \triangleq \boldsymbol{\Phi x} + \boldsymbol{\eta}$ (where $\boldsymbol{\Phi}$ is a $m \times n$ sensing matrix and $\boldsymbol{\eta}$ is a noise vector with $m$ elements), with rigorous theoretical guarantees at even sub-Nyquist sampling rates \cite{Donoho_2006}\cite{Candes_2008}. The quality of the reconstruction depends on the choice of the sensing matrix $\boldsymbol{\Phi}$ and on how sparse the signal is in $\boldsymbol{\Psi}$. Successful recovery with theoretical guarantees is possible with efficient algorithms like Basis pursuit which minimizes the $\ell_{1}$ norm of the sparse coefficient vector ($\boldsymbol{\theta}$) such that $\|\boldsymbol{y}-\boldsymbol{\Phi\Psi\theta}\|_{2} \leq \epsilon$ where $\|\boldsymbol{\eta}\|_2 < \epsilon$ \cite{Candes_2008_ric}. One well-studied sufficient condition on $\boldsymbol{\Phi}$ for guaranteeing good quality reconstructions is the Restricted Isometric Property (RIP) \cite{Candes_2005}. It can also be shown that matrices drawn from Gaussian or Bernoulli distributions satisfy RIP with high probability \cite{Candes_2008_ric}.

At the same time, it has been shown that optimizing $\boldsymbol{\Phi}$ w.r.t. a carefully designed `matrix quality measure' leads to better-than-random recovery. \cite{Elad_2007,Duarte_2009,Duarte_2012,Mangia_2018,Abol_2010,Arce_2014,Zickler_2011,Bald_2016,Gozcu_2018}. However, computing the Restricted Isometric Constant (RIC), which characterizes the RIP, is known to be computationally prohibitive. Hence, other measures have been proposed, such as mutual coherence \cite{Candes_2005} or average mutual coherence as they are computationally much more efficient. \cite{Cai_2010} also shows that improving the mutual coherence improves a worst case bound on the reconstruction error. The mutual coherence is defined as $\mu_{max} \triangleq \|\boldsymbol{A}^{T}\boldsymbol{A}-\boldsymbol{I}\|_{\infty}$ wherein the columns of $\boldsymbol{A} \triangleq \boldsymbol{\Phi \Psi}$ are taken to be unit-normalized. Further, to relax the $\ell_{\infty}$ norm, approaches such as minimizing the average mutual coherence $\mu_{avg} \triangleq \|\boldsymbol{A}^{T}\boldsymbol{A}-\boldsymbol{I}\|_{F}$ have also been proposed \cite{Duarte_2009, Duarte_2012, Abol_2010}.

To further enhance the quality of reconstruction, some approaches have also focused on incorporating prior information while designing the sensing matrix. For example, in the work proposed in \cite{Li_2017}, a weighted average mutual coherence minimization is proposed. The approach is used to reconstruct video frames wherein the the recovered image from the previous frame measurement is used to create weights for the subsequent frame to assign relative importance values to different columns of the matrix $\boldsymbol{A}$. A piecewise-linear estimator (PLE) with a Gaussian Mixture Model (GMM) prior was introduced in \cite{Yu_2012}, which was extended to the Statistical Compressive Sensing (SCS) framework \cite{Sapiro_2011}. \cite{Shah_2018} evolved a design method based on the minimum mean squared error (MMSE) criterion, imposing priors on signal (for natural images) and using noise negative log-likelihood.

While most of the methods in literature have focused on optimizing the sensing matrix based on certain bounds, \cite{Bald_2016} provides an approach to subsample from the signal/image via the identification of modularity and submodularity structures. Besides Restricted Isometric Property and mutual coherence, the $\ell_1-\ell_{\infty}$ measure was proposed in \cite{Tang_2015}. The $\ell_1-\ell_{\infty}$ measure provides tighter reconstruction bounds than the coherence while being computationally easier to calculate than the Restricted Isometric Constant. It is defined as $\omega_2(\boldsymbol{Q},s) \triangleq \textrm{min}_{\boldsymbol{z}: \|\boldsymbol{z}\|_1/\|\boldsymbol{z}\|_{\infty} \leq s} \|\boldsymbol{Qz}\|_2/\|\boldsymbol{z}\|_{\infty}$ where $\boldsymbol{Q} \triangleq \boldsymbol{A}^T\boldsymbol{A}$ and $s$ is a sparsity-parameter. 

In this paper we propose a generalization of the mutual coherence criterion by proposing our bi-coherence and tri-coherence approaches. Our proposed criteria are easier to optimize on, as compared to \cite{Tang_2015} as well as the RIC. We also take into account training data to further improve our sensing matrices through our proposed weighted coherence, weighted bi-coherence and weighted tri-coherence criteria. Finally, we propose algorithms to solve the proposed maximization problems starting with random binary or random $U(0,1)$ sensing matrices.
\section{Acquisition Model}
\label{sec:format}
We consider designing sensing matrices for the block single pixel camera acquisition model as described in \cite{Kerviche_2014}. In this architecture, compressive measurements are acquired independently for each non-overlapping image block of size $n \times n$. Consider there are ($m \ll n^2$) measurements available per block. Thus, our measurements $\boldsymbol{y_{i}}$ corresponding to the $i^{\textrm{th}}$ block $\boldsymbol{x_{i}}$ can be written as $\boldsymbol{y_{i}} = \boldsymbol{\Phi} \boldsymbol{x_{i}} + \boldsymbol{\eta_{i}}$ where $\boldsymbol{x_{i}}$ denotes the $i^{\textrm{th}}$ vectorized block and $\boldsymbol{\eta_{i}}$ denotes the measurement noise vector containing elements drawn i.i.d. from $\mathcal{N}(0,\sigma^2)$. Furthermore, we consider that the elements of the sensing matrix $\boldsymbol{\Phi}$ are drawn from either a Bernoulli $\{0,1\}$ or a $U(0,1)$ distribution. 

\section{Proposed Approach}
\label{sec:pagestyle}
We propose generalized coherence based approaches for matrix design. Coherence is a matrix quality measure that discourages any one column $\boldsymbol{A_i}$ of the matrix from being too similar to any other column $\boldsymbol{A_j}$ \cite[Fig. 1, Page 3]{Romberg2008}. The main intuition behind our
approach is to discourage any column vector of the sensing matrix from being accurately expressible as a \emph{sparse linear combination} of other column vectors. Optimizing on such a generalized coherence criterion ensures that during sparse recovery, the picking of the non-zero atoms becomes easier. This is because the sensing matrix column corresponding to that atom would now be hard to replace using other column vectors. Based on this idea, we propose bi-coherence and tri-coherence approaches.
\subsection{Bi-coherence}
The bi-coherence criterion is based on the idea that any column vector of the matrix $\boldsymbol{A} = \boldsymbol{\Phi \Psi}$ should not be able to be written as a linear combination of any other two column vectors of $\boldsymbol{A}$. Thus, the following loss function results:
\begin{align}
& \boldsymbol{A} = \textrm{argmax}_{\boldsymbol{A}} \sum_{i,j>i,k>j}^{n}  \textrm{argmin}_{\alpha(i,j,k)} \| \boldsymbol{A_{i}}-\boldsymbol{A_{j,k}}\alpha(i,j,k) \|_{2}^2,
\end{align}
where $\boldsymbol{A_{i}}$ denotes the $i^{\textrm{th}}$ column of $\boldsymbol{A}$, $\boldsymbol{A_{j,k}}$ is the $n \times 2$ matrix formed by concatenating the $j^{\textrm{th}}$ and $k^{\textrm{th}}$ columns of $\boldsymbol{A}$, and $\alpha(i,j,k)$ is a $2 \times 1$ coefficient vector. The minimization part can be solved by assigning $\alpha(i,j,k) = (\boldsymbol{A_{j,k}}^{T} \boldsymbol{A_{j,k}})^{-1} \boldsymbol{A_{j,k}}^{T} \boldsymbol{A_{i}}$. 
Thus, the following optimization problem results:
\begin{align}
\boldsymbol{A} = \textrm{argmax}_{\boldsymbol{A}}  \sum_{i,j>i,k>j}^{} 
\| \boldsymbol{A_{i}}-\boldsymbol{A_{j,k}}(\boldsymbol{A_{j,k}}^{T} \boldsymbol{A_{j,k}})^{-1} \boldsymbol{A_{j,k}}^{T} \boldsymbol{A_{i}} \|_{2}^2.  
\end{align}
\subsection{Tri-coherence}
Similar to bi-coherence, the tri-coherence criterion would discourage that any column vector of the matrix $\boldsymbol{A}$ from being accurately expressible as a linear combination of any other three column vectors of $\boldsymbol{A}$, giving rise to the following problem:
\begin{align}
\boldsymbol{A} = \textrm{argmax}_{\boldsymbol{A}} \sum_{i,j>i,k>j,l>k}^{} \nonumber \textrm{argmin}_{\alpha(i,j,k,l)} \| \boldsymbol{A_{i}}- \boldsymbol{A_{j,k,l}}\alpha(i,j,k,l)\|_{2}^2,
\end{align}
where $\boldsymbol{A_{j,k,l}}$ is the $n \times 3$ matrix formed by concatenating the $j^{\textrm{th}}$, $k^{\textrm{th}}$ and $l^{\textrm{th}}$ columns of $\boldsymbol{A}$. Following steps similar to bi-coherence, we get
\begin{align}
\boldsymbol{A} = \textrm{argmax}_{\boldsymbol{A}}\sum_{i,j>i,k>j,l>k}^{} \| \boldsymbol{A_{i}}- \boldsymbol{A_{j,k,l}}(\boldsymbol{A_{j,k,l}}^{T} \boldsymbol{A_{j,k,l}})^{-1} \boldsymbol{A_{j,k,l}}^{T} \boldsymbol{A_{i}}\|_{2}^2.
\end{align}
\subsection{Relationship with the average coherence minimization criteria}
Similar to bi-coherence and tri-coherence, if we have a single column vector $\boldsymbol{A}_{j}$, we will have the following loss function,
\begin{align}
 \boldsymbol{A} =& \textrm{argmax}_{\boldsymbol{A}} \sum_{i,j>i}^{}
     \textrm{argmin}_{\alpha(i,j)} \| \boldsymbol{A_{i}}- \boldsymbol{A_{j}}\alpha(i,j)\|_{2}^2 \nonumber \\
     =& \textrm{argmax}_{\boldsymbol{A}} \sum_{i,j>i}^{} \| \boldsymbol{A_{i}}-\boldsymbol{A_{j}}(\boldsymbol{A_{j}}^{T} \boldsymbol{A_{j}})^{-1} \boldsymbol{A_{j}}^{T} \boldsymbol{A_{i}} \|_{2}^2  \nonumber \\
     =& \textrm{argmax}_{\boldsymbol{A}} \sum_{i,j>i}^{} \|\boldsymbol{A_{i}}\|^{2} - \frac{(\boldsymbol{A_{j}}^{T}\boldsymbol{A_{i}})^2}{ \|\boldsymbol{A_{j}}\|^{2}}
\end{align}
For the case of unit normalized columns of $\boldsymbol{A}$, the resulting expression is same as minimizing average coherence.
\subsection{Weights based approach using training data}
We also propose weighted coherence based approaches for cases in which training data is available. The main idea is that certain frequencies in the DCT or wavelet basis contribute more to the signal than others. Hence, we assign weights to emphasize certain groups of column vectors over other column vectors. For this, we compute the mean absolute value of the coefficients at the $l^{\textrm{th}}$ frequency (in a chosen $\boldsymbol{\Psi}$) over a training set, and set it to be equal to the weight $W_l$. Thus, $\boldsymbol{W}$ is a vector of weights where each element represents the weight assigned to a certain frequency. Using the weights we propose weighted coherence, weighted bi-coherence and weighted tri-coherence.
\\
1) \textbf{Weighted Coherence}
\begin{align}
\boldsymbol{A} = \textrm{argmax}_{\boldsymbol{A}} \sum_{i,j>i}^{n} \textrm{argmin}_{\alpha(i,j)} \sqrt{\boldsymbol{W}(i)\cdot \boldsymbol{W}(j)}\|\boldsymbol{A_{i}}-\boldsymbol{A_{j}}\alpha(i,j)\|_{2}^2
\end{align}
where $\boldsymbol{W}$ is the weight vector obtained from training data. Following steps similar to the unweighted approaches, we get
\begin{align}
\boldsymbol{A} = \textrm{argmax}_{\boldsymbol{A}} \sum_{i,j>i}^{n}   \sqrt{\boldsymbol{W}(i)\cdot \boldsymbol{W}(j)}\|\boldsymbol{A_{i}}-\boldsymbol{A_{j}}(\boldsymbol{A_{j}}^{T} \boldsymbol{A_{j}})^{-1} \boldsymbol{A_{j}}^{T} \boldsymbol{A_{i}} \|_{2}^2.
\end{align}
2) \textbf{Weighted bi-Coherence}
\begin{align}
\boldsymbol{A} = \textrm{argmax}_{\boldsymbol{A}}  \sum_{i,j>i,k>j}^{} (\boldsymbol{W}(i)\cdot \boldsymbol{W}(j)\cdot \boldsymbol{W}(k))^\frac{1}{3}\nonumber\\ \|\boldsymbol{A_{i}}-\boldsymbol{A_{j,k}}(\boldsymbol{A_{j,k}}^{T} \boldsymbol{A_{j,k}})^{-1} \boldsymbol{A_{j,k}}^{T} \boldsymbol{A_{i}} \|_{2}^2.
\end{align}
3) \textbf{Weighted tri-Coherence}
\begin{align}
    & \boldsymbol{A} = \textrm{argmax}_{\boldsymbol{A}} \sum_{i,j>i,k>j,l>k}^{}  \nonumber \\ 
    & (\boldsymbol{W}(i)\cdot \boldsymbol{W}(j)\cdot \boldsymbol{W}(k) \cdot \boldsymbol{W}(l))^\frac{1}{4} \nonumber \\
    & \|\boldsymbol{A_{i}}-\boldsymbol{A_{j,k,l}}(\boldsymbol{A_{j,k,l}}^{T} \boldsymbol{A_{j,k,l}})^{-1} \boldsymbol{A_{j,k,l}}^{T} \boldsymbol{A_{i}} \|_{2}^2.
\end{align}
\subsection{Optimizing over the matrices}
For the binary matrices, we propose Alg. \ref{alg:algorithm1}. Here $L(.)$ denotes the loss function. We iterate over the entire sensing matrix and check whether switching the binary value at any entry improves the loss at every location of the matrix. This is repeated for several epochs. Convergence is achieved when the increase in the loss function for successive epochs is less than some convergence parameter $\delta$. For optimizing over continuous-valued matrices (starting with a random $U(0,1)$ initial condition), we use projected gradient ascent. While updating the matrix, we first perform a gradient ascent step and then project the resultant matrix on the space of matrices (unit normalized columns) with entries confined to $[0,1]$. 
\begin{algorithm}
\SetAlgoLined
 take $\boldsymbol{\phi}$ as random binary matrix\;
 \While{convergence}{
 \For{$i = 1:m$}{
 \For{$j = 1:n$}{
 $l_{1} = L(\boldsymbol{\phi})$\;
 $\boldsymbol{\phi}(i,j) = 1-\boldsymbol{\phi}(i,j)$\;
  $l_{2} = L(\boldsymbol{\phi})$\;
  \If{$l_{2}\leq l_{1}$}{
 $\boldsymbol{\phi}(i,j) = 1-\boldsymbol{\phi}(i,j)$\;
   }
 }
 }
 }
\caption{Binary matrix optimization}
\label{alg:algorithm1}
\end{algorithm}
\section{Evaluations}
We perform reconstructions on the compressive measurements from cropped images of the Berkley dataset (image size = $160\times240$). The patch size is taken to be $10\times10$ and iid noise from $\mathcal{N}(0,4)$ is added to the compressive measurements. The DCT basis is used for sparse representation. A block-based sparse reconstruction is preferred instead of the Rice single pixel approach \cite{Duarte_2008} owing to the large image size. Reconstruction is performed on 100 test images while the weights are generated using 200 training images (with no overlap with the test images). The convergence parameter $\delta$ is chosen to be 0.001. The basis pursuit algorithm is used for reconstruction using the SPGL solver \cite{BergFriedlander_2008, spgl1site}.
with the noise parameter $\epsilon \geq \|\boldsymbol{\eta}\|_2$ chosen using cross-validation.
\subsection{Results for continuous-valued sensing matrices} 
\begin{figure}[H]
    \centering
    \includegraphics[scale=0.13]{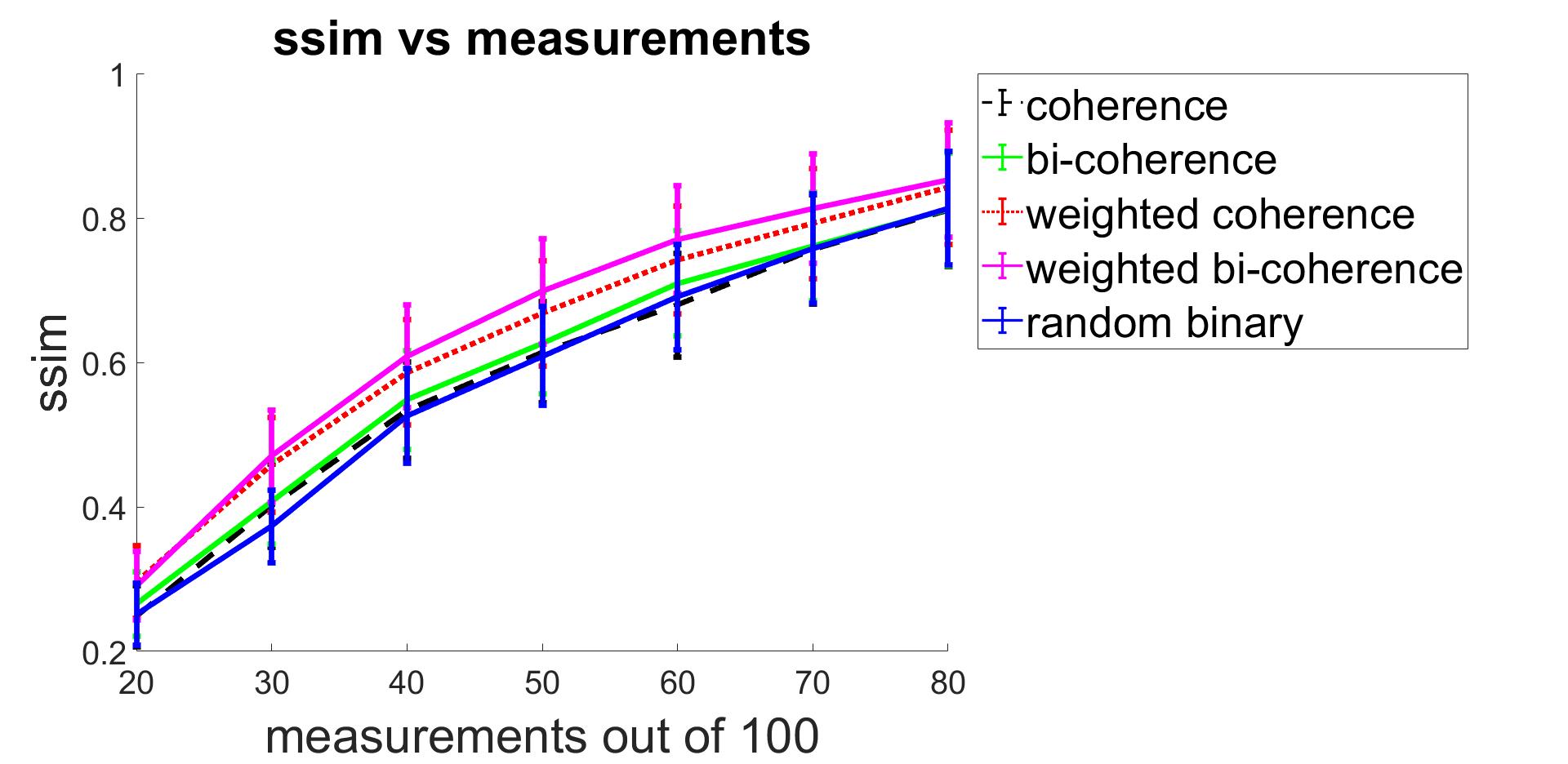}
        \caption{SSIM comparison of different coherence based methods on continuous-valued matrices}
        \label{fig:ssim_uniform}
\end{figure}

\begin{figure}[H]
    \centering
    \includegraphics[scale=0.47]{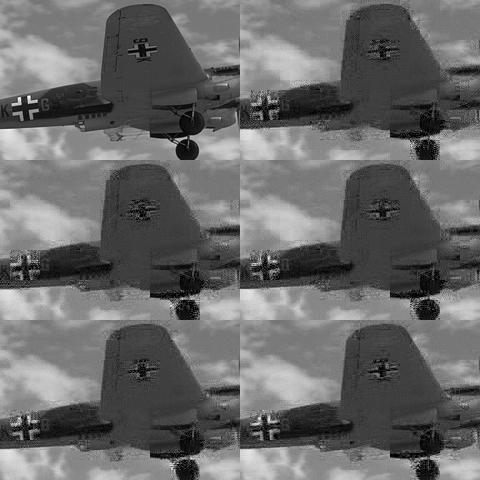}
            \caption{$m=40, n^2=100$; Left to right top to bottom: ground truth image, reconstructions with random uniform (PSNR 27.31), coherence (27.46), double coherence (28), weighted coherence (28.93) and weighted double coherence (29.25).}
\label{fig:unif1}
\end{figure}

\begin{figure}[H]
    \centering
    \includegraphics[scale=0.47]{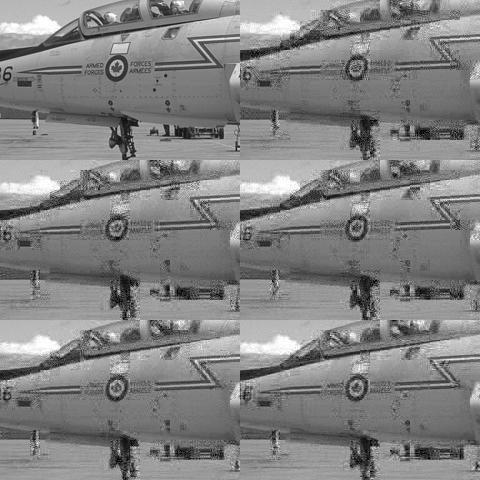}
            \caption{$m = 40, n^2 = 100$; Left to right top to bottom: ground truth image, reconstructions with random uniform (PSNR 24.42), coherence (24.61), double coherence (25.06), weighted coherence (25.84) and weighted double coherence (26.24).}
\label{fig:unif2}
\end{figure}

\begin{figure}[H]
    \centering
    \includegraphics[scale=0.47]{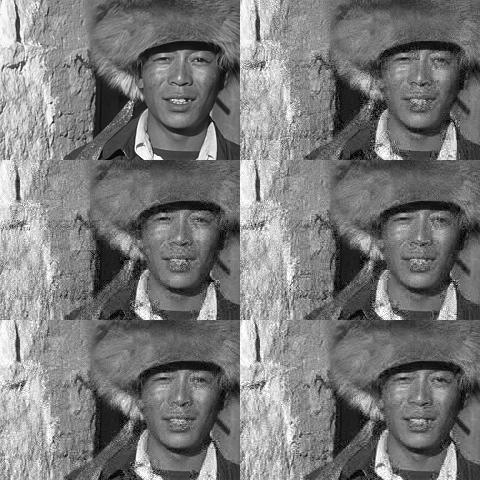}
            \caption{$m = 50, n^2 = 100$; Left to right top to bottom: ground truth image, reconstructions with random uniform (PSNR 23.39), coherence (23.51), double coherence (23.9), weighted coherence (24.78) and weighted double coherence (25.31).}
\label{fig:unif3}
\end{figure}

\begin{figure}[H]
    \centering
    \includegraphics[scale=0.47]{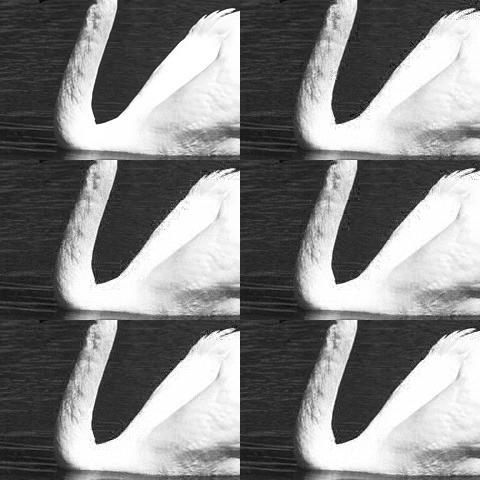}
            \caption{$m = 70, n^2 = 100$;
            Left to right top to bottom: ground truth image, reconstructions with random uniform (PSNR 32.05), coherence (32.14), double coherence (32.57), weighted coherence (33.55) and weighted double coherence (34.77).}
\label{fig:unif4}
\end{figure}

Here, we present results with continuous-valued sensing matrices optimized based on our proposed coherence, bi-coherence, weighted coherence and weighted bi-coherence criteria. We can see from the SSIM values in Fig. \ref{fig:ssim_uniform} that the weighted versions perform better than the ones without training data with weighted bi-coherence performing the best. Reconstructed images are presented in Figs. \ref{fig:unif1}, \ref{fig:unif2}, \ref{fig:unif3} and \ref{fig:unif4}. The results with weighted bicoherence typically show less block artifacts and better edge reconstruction than with others.

\subsection{Comparison for binary matrices}
\begin{figure}[H]
\centering
    \includegraphics[scale=0.13]{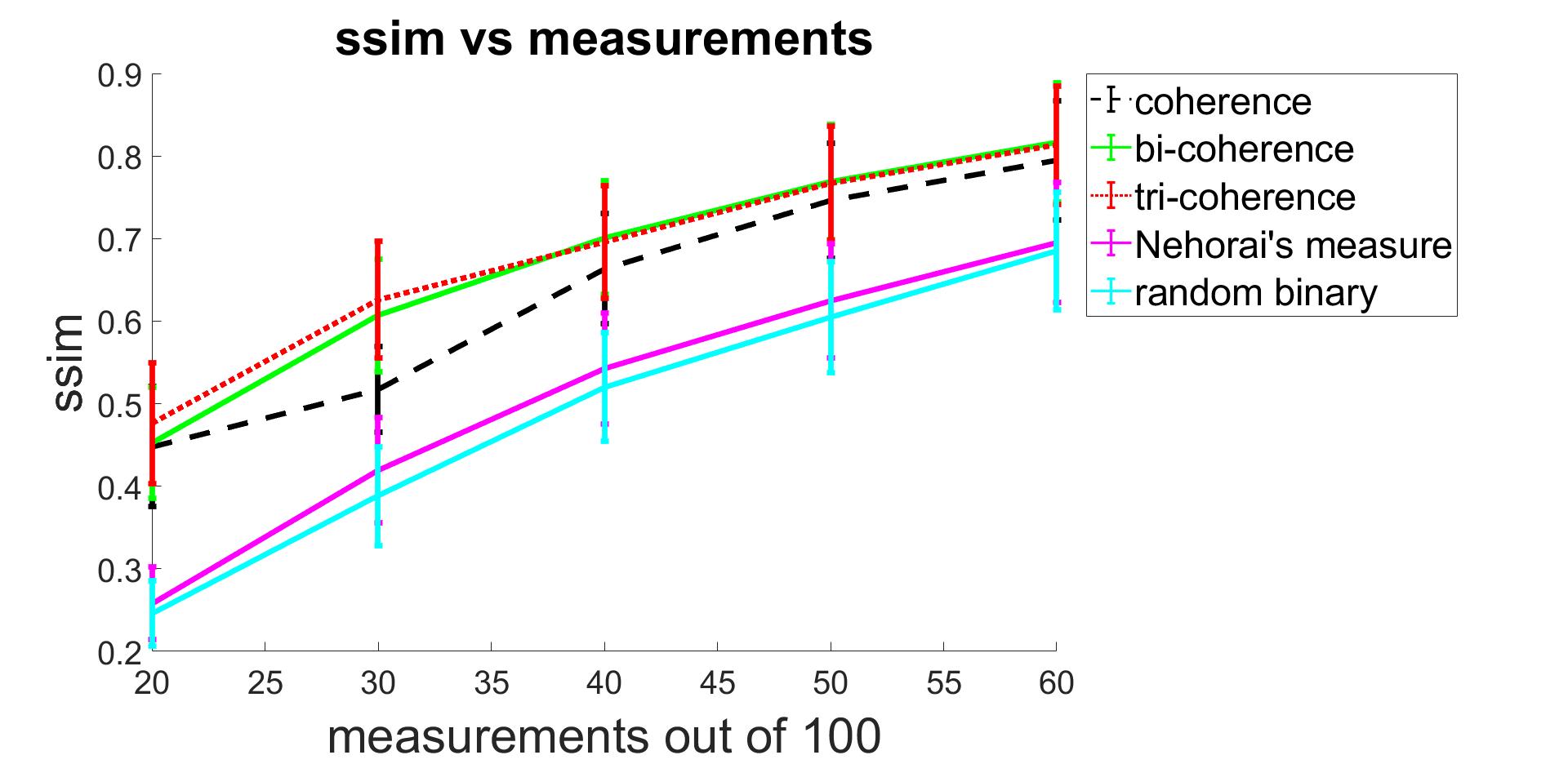}
        \caption{SSIM comparison of different non-data driven matrix optimization methods on binary matrices}
        \label{fig:ssim_nondatadriven}
\end{figure}

\begin{figure}[H]
    \centering
    \includegraphics[scale=0.47]{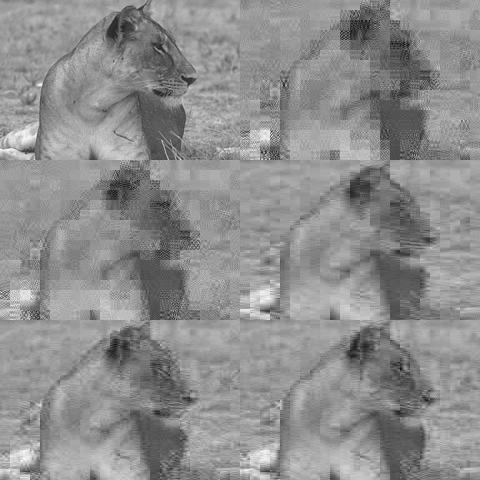}
            \caption{$m = 20, n^2 = 100$;
            Left to right top to bottom: ground truth image, reconstructions with random binary (PSNR 22.65), $l_{1}-l_{\infty}$ (23.47), coherence (26.71), double coherence (26.77) and triple coherence (27.3).}
            \label{fig:binary1}
\end{figure}

\begin{figure}[H]
    \centering
    \includegraphics[scale=0.47]{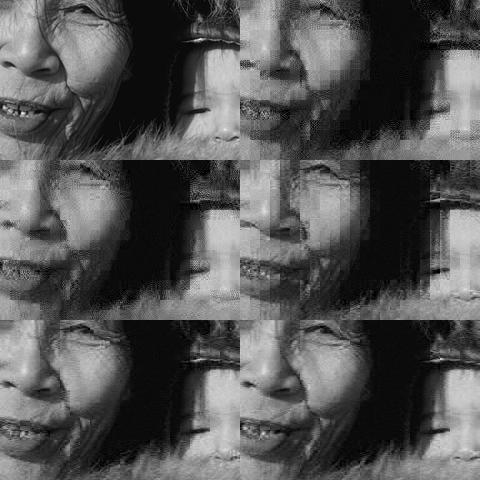}
            \caption{$m = 30, n^2 = 100$;
            Left to right top to bottom: ground truth image, reconstructions with random binary (PSNR 23.42), $l_{1}-l_{\infty}$ (25.41), coherence (27.68), double coherence (29.76) and triple coherence (29.21).}
\label{fig:binary2}
\end{figure}

\begin{figure}[H]
    \centering
    \includegraphics[scale=0.47]{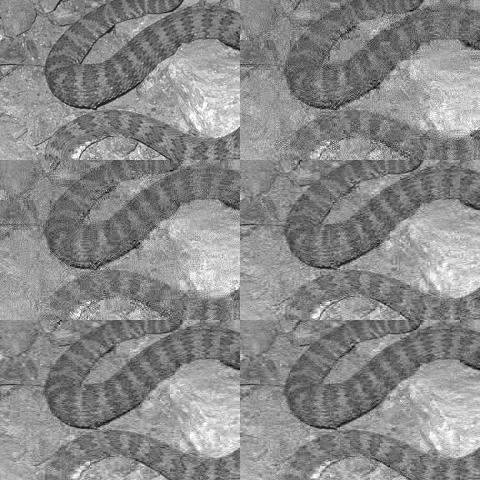}
            \caption{$m = 40, n^2 = 100$; Left to right, top to bottom: ground truth image, reconstructions with random binary (PSNR 22.53), $l_{1}-l_{\infty}$ approach (23), coherence (25.02), double coherence (25.73) and triple coherence (25.77).}
\label{fig:binary3}
\end{figure}

\begin{figure}[H]
    \centering
    \includegraphics[scale=0.47]{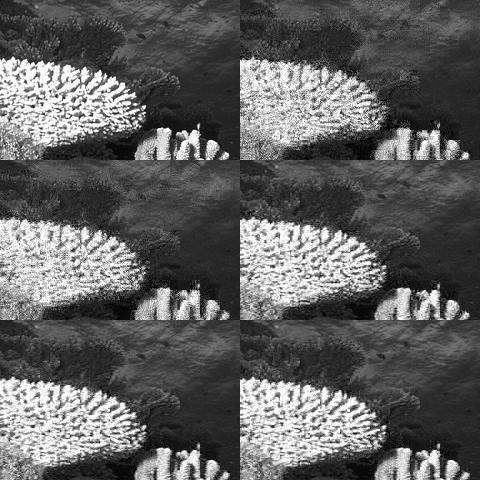}
            \caption{$m = 50, n^2 = 100$;
            Left to right, top to bottom: ground truth image, reconstructions with random binary (PSNR 20.37), $l_{1}-l_{\infty}$ approach (20.95), coherence (23.47), double coherence (24.44) and triple coherence (24.44).}
\label{fig:binary4}
\end{figure}

\begin{figure}[H]
    \centering
    \includegraphics[scale=0.47]{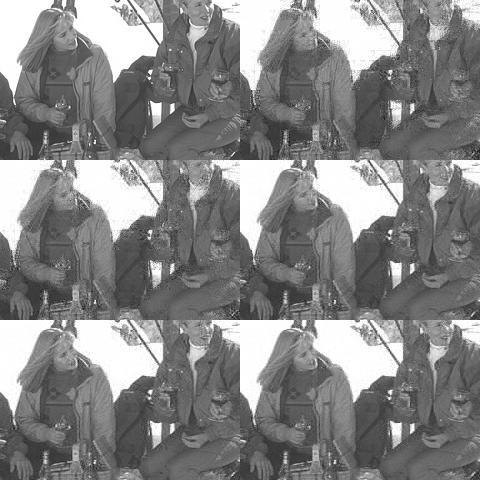}
            \caption{$m = 60, n^2 = 100$;
            Left to right, top to bottom: ground truth image, reconstructions with random binary (PSNR 26.65), $l_{1}-l_{\infty}$ approach (26.96), coherence (29.65), double coherence (30.19) and triple coherence (29.94).}
\label{fig:binary5}
\end{figure}

In this section, we first present results with non data-driven techniques such as by maximizing coherence, bi-coherence, tri-coherence and the $\ell_1-\ell_{\infty}$ measure from \cite{Tang_2015}. The hyperparameter $s$ for the $\ell_1-\ell_{\infty}$ measure is chosen empirically, by observing the range of values of $\|\boldsymbol{\theta}\|_1/\|\boldsymbol{\theta}\|_{\infty}$ on a training set of image patches (here $\boldsymbol{\theta}$ is the vector of transform coefficients of a patch). The average is observed to be 2 while the 99 percentile is around 5. Hence, we choose the best $s$ within this range. The average SSIM values are plotted in Fig. \ref{fig:ssim_nondatadriven} and image results are presented in Figs. \ref{fig:binary1}, \ref{fig:binary2}, \ref{fig:binary3}, \ref{fig:binary4} and \ref{fig:binary5}, showing best results with bi-coherence or tri-coherence.

\begin{figure}[H]
    \centering
    \includegraphics[scale=0.13]{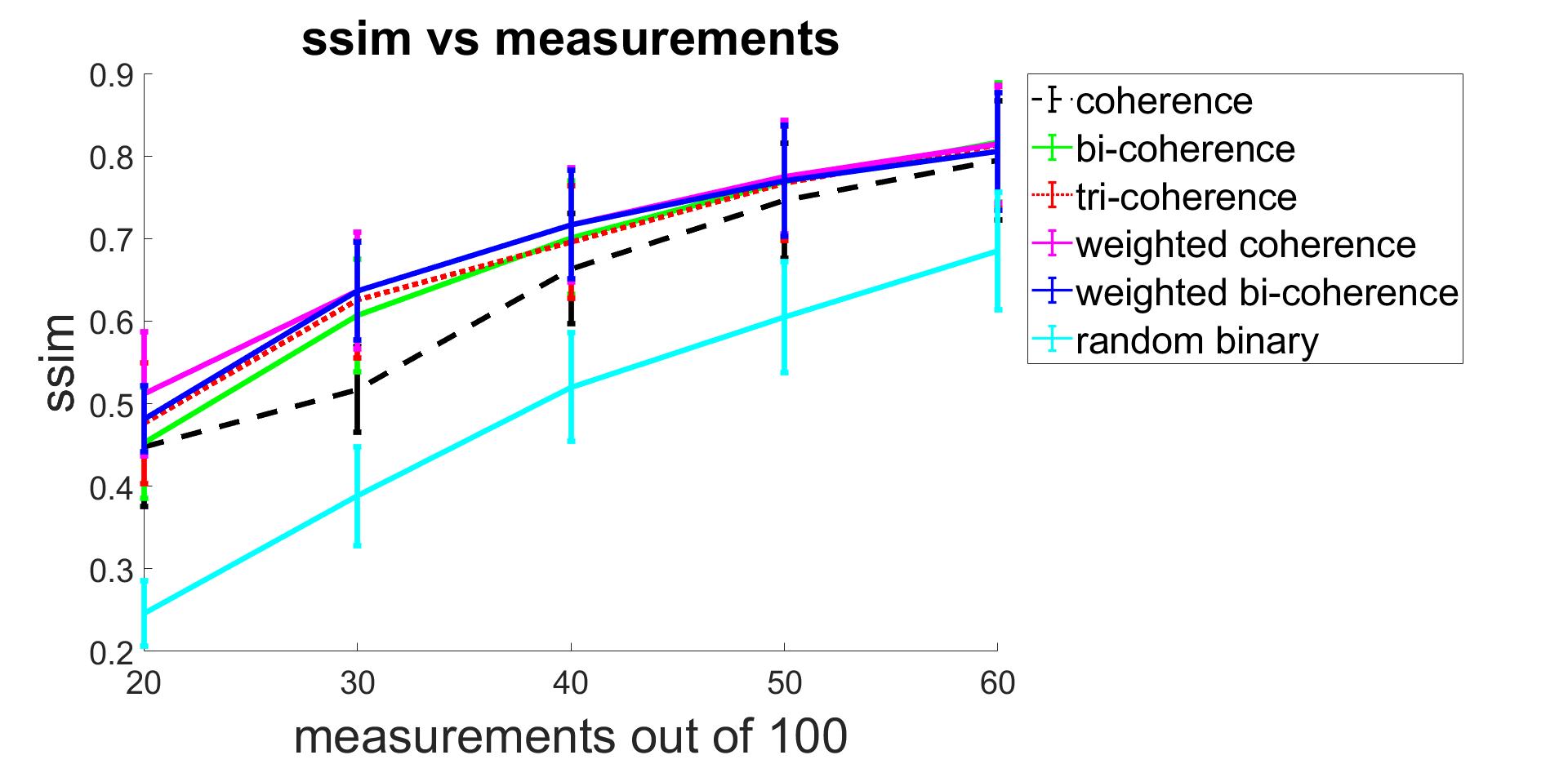}
        \caption{SSIM comparison of different coherence based methods on binary matrices}
        \label{fig:ssim_coherence}
\end{figure}

\begin{figure}[H]
    \centering
    \includegraphics[scale=0.35]{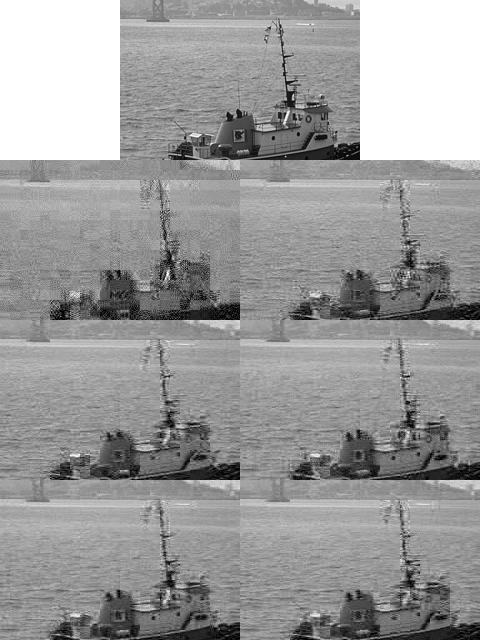}
            \caption{$m = 30, n^2 = 100$; From left to right, top to bottom: ground truth image, reconstruction with random binary (21.73), coherence (24.79), double coherence (25.57), triple coherence (25.94), weighted coherence (26.2) and weighted double coherence (26.41).}
\label{fig:coh1}
\end{figure}

\begin{figure}[H]
    \centering
    \includegraphics[scale=0.35]{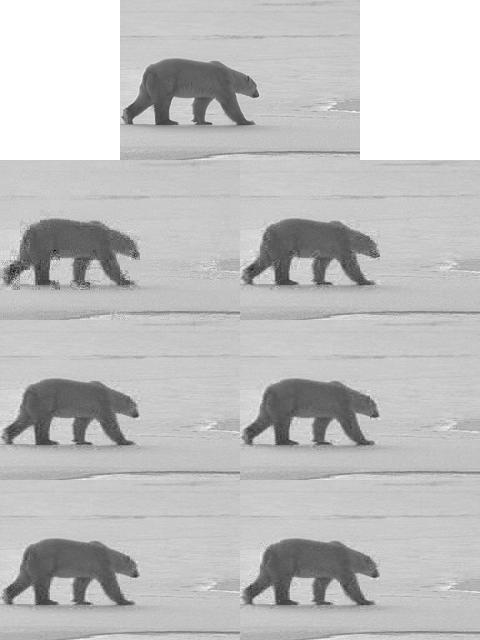}
            \caption{$m = 40, n^2 = 100$; Left to right, top to bottom: ground truth image, reconstructions with random binary (PSNR 30.21), coherence (33.78), double coherence (35.22), triple coherence (35), weighted coherence (35.33) and weighted double coherence (35.52).}
\label{fig:coh2}
\end{figure}

\begin{figure}[H]
    \centering
    \includegraphics[scale=0.35]{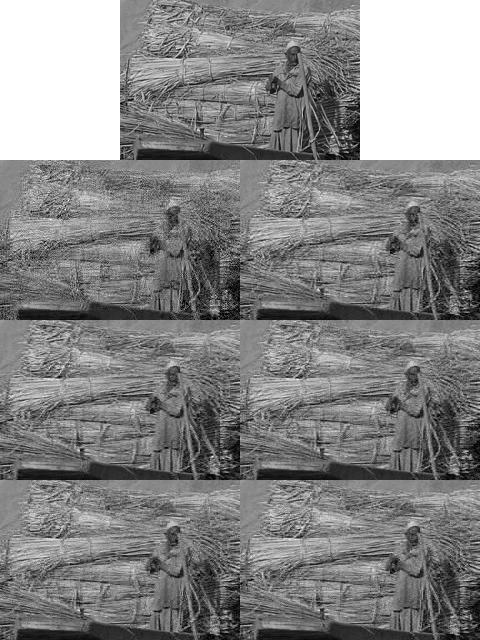}
            \caption{$m = 50, n^2 = 100$; Left to right, top to bottom: ground truth image, reconstructions with random binary (PSNR 21.52), coherence (24.61), double coherence (25.35), triple coherence (25.31), weighted coherence (25.26) and weighted double coherence (25.2).}
\label{fig:coh3}
\end{figure}

\begin{figure}[H]
    \centering
    \includegraphics[scale=0.35]{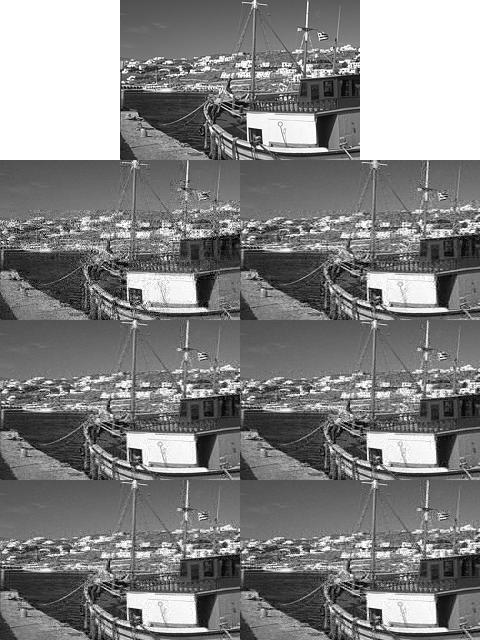}
            \caption{$m = 60, n^2 = 100$; Left to right, top to bottom: ground truth image, reconstructions with random binary (PSNR 22.08), coherence (24.95), double coherence (25.83), triple coherence (25.64), weighted coherence (25.82) and weighted double coherence (25.6).}
\label{fig:coh4}
\end{figure}

We now present results on binary matrices, optimized by maximizing coherence, double coherence, triple coherence, weighted coherence and weighted double coherence. The SSIM values are plotted in Fig. \ref{fig:ssim_coherence}. We can see that the weighted versions perform better than the ones without training data. A few image results are presented in Figs. \ref{fig:coh1}, \ref{fig:coh2}, \ref{fig:coh3} and \ref{fig:coh4}.

\section{Conclusion and Future work}
\label{sec:print}
In this paper, we have proposed a new set of quality measures for sensing matrix design, showing improved quality of reconstruction compared to random matrices or optimizing over measures such as the average coherence of the $\ell_1-\ell_{\infty}$ measure. Furthermore, the expressions for our quality measures are available in closed form. Hence, these measures can be optimized using gradient-based methods, which is not easily possible in case of the $\ell_1-\ell_{\infty}$ measure. Likewise, the RIC is also computationally expensive as it requires computing singular values of many sub-matrices and is difficult to optimize over, due to lack of availability of closed-form expressions. One could argue that our proposed design criterion could be computationally expensive if we move further from tri-coherence (i.e. increase the number of column vectors for the linear combination). However, even for the $\ell_1-\ell_{\infty}$ measure, the optimal hyperparameter $s-1$ was found to be around 3 to 4, for image patch reconstruction problems. We have also modified our quality measures to the case where training data are available, producing weighted quality measures. Besides potentially improving encoder design in practical imaging systems, another direct application of our work could be in the field of encoder design for group testing \cite{Ghosh_2020}. Exploring theoretical performance guarantees for the proposed measures is another avenue for future research. 

\bibliographystyle{IEEEbib}
\bibliography{refs}

\end{document}